\documentclass[10pt,a4paper]{article} 
\usepackage{lipsum}
\usepackage{url}
\usepackage[nochapters]{classicthesis} 

\usepackage{amsthm}
\usepackage{amsfonts}
\usepackage{amsmath}
\usepackage{amssymb}
\usepackage{nicefrac}
\usepackage{graphicx}
\usepackage{booktabs}
\usepackage{footnote}
\usepackage[english]{babel}
\usepackage[T1]{fontenc} 
\usepackage{microtype}
\usepackage{afterpage}
\usepackage[affil-it]{authblk}
\newcounter{thmcount}
\theoremstyle{remark}
\newcommand{\citep}[1]{\cite{#1}}

\begin{document}
    \title{\rmfamily\normalfont\spacedallcaps{On the statistical
        description of the inbound air traffic over Heatrow Airport}} 

    \author[1]{Maria Virginia Caccavale}
    \author[1]{Antonio Iovanella}
    \author[1,3]{Carlo Lancia}
    \author[2]{Guglielmo Lulli}
    \author[1]{Benedetto Scoppola}
    \affil[1]{Universit\`a di Roma `Tor Vergata'}
    \affil[2]{Universit\`a Milano `Bicocca'}
    \affil[3]{TU/e, Technische Universiteit Eindhoven}

    \maketitle

    \begin{abstract}
    We present a model to describe the inbound air traffic over a congested hub.
    We show that this model gives a very accurate description of the traffic by the
    comparison of our theoretical distribution of the queue with the
    actual distribution observed over Heathrow airport. 
    We discuss also the robustness of our model.
    
    \end{abstract}
       
    
    \section{Introduction}
\label{sec:introduction}

Airport congestion is a persistent phenomena in air traffic.  Air traffic congestion is significant
even if the principal airports in Western and Central Europe are
treated as ``fully coordinated''\footnote{In the U.S., scheduling
  limits are applied only to New York 
  region airports, Washington/Reagan, and Chicago/O'Hare airport,
  under the High Density Rule (HDR).},  
meaning essentially that the number of flights that can be
scheduled there per hour (or other unit of time) is not allowed to
exceed the ``declared capacity'' of the airports~\citep{deNO2003}. In
2011,  the average additional  Arrival Sequencing and Metering
Area (ASMA) time\footnote{ The Arrival Sequencing and Metering Area (ASMA) is
  the airspace within a radius of 40NM around an airport. The ASMA
  additional time is a proxy for the average arrival runway queuing
  time on the inbound traffic flow, during times when the airport is
  congested.} 
at the top 30 european airports amounted to  2.9 minutes per arrival increasing
by +5\% with respect to the previous year. On this statistic, London
Heathrow is a clear outlier, having by far the highest level of
additional time within the last 40NM with 8.2 minutes per arrival,
followed by Frankfurt and Madrid~\citep{PRR}. Similar situations occur
in the US~\citep{ball1}.
\newline
Quoting the 2011 Performance Review  Report:  ``Airports are key
nodes of the aviation network and airport capacity is considered to be
one of the main challenges to future air traffic growth. This requires
an increased focus on the integration of airports in the ATM network
and the optimisation of operations at and around airports''.  
\newline
Several approaches have been proposed to  mitigate congestion and
resolve demand-capacity imbalances.  At operational level
(short-term), these  approaches consider the operational adjustment of
air traffic flows to match available capacity. So far, the most popular
approach in resolving these short-term periods of congestion
has shown to be the allocation of ground delays~\citep{Od1987}. The Ground
Holding Problem considers the development of strategies for allocating
ground delays to aircraft, and has received considerable
attention~\citep{RO1994, DL2003, BHM2010, ADL2011}.
However, these air traffic 
flow management strategies might be suboptimal because they do not
capture the inherent unpredictability  of arrivals at
airports. \cite{ball1} showed that  changes in the current
practice 
for setting airport arrival rates can lead to significant benefits in
terms of additional ASMA times. 
\newline
In view of the current situation, it is extremely important to have a reliable tool to measure and  forecast congestion in the Air Traffic System. However, in developing such a tool there are some issues to address. First of all, the  stochastic models developed so far to describe air traffic congestion are not reliable.  
The interarrival times between two consecutive arrivals at an airport seem to be distributed exponentially, leading to the hypothesis that the arrivals are Poissonian  ~\citep{willemain2004statistical}. Nonetheless, the  results of the classical queueing theory with Poisson arrivals do not fit with the observed data; see Figure~\ref{fig:heathrow-fit} below. 

A second issue regards the validation of the stochastic processes. In fact, it  is not easy to  quantify the amount of congestion, because it is difficult to evaluate the number of aircrafts in queue.  

In this paper we propose a straightforward description of  the queueing
process at a very congested airport. We study the inbound air traffic at Heathrow airport.
We compute the time spent in queue calculating the difference between
the actual and the minimal time spent in the vicinities of the
airport. The extra time is considered to be time spent in queue (the
details of this simple computation are in Section~\ref{sec:heathrow}).  
This procedure solves the problem of evaluating the distribution of the queue in this very congested case.

We present a quite natural mathematical model for
the arrival process of the aircraft, that is very different from the Poisson
process but has a distribution of the interarrival times very close to the
exponential.
The idea of the process, that will be described in details in the next section,
is easy: we start with a deterministic schedule, organized in such a way
that the aircraft should arrive in a deterministic and homogeneous way
(say with a constant interarrival time similar to the time needed for a landing),
and then we add an independent random variable to each of this scheduled
arrivals. The resulting list of times of arrival is then mixed up by the random
delay. The process obtained in this way has a long history~\citep{Kendall1964}. 
It is easy to study numerically but quite difficult to treat from a
mathematical point of view, though significative progresses have
been recently made~\citep{gns,eda-2012}.

The simulations we present in this paper show the following facts:
\begin{enumerate}
\item The fit with the real data over Heathrow airport is really excellent,
and it is incomparably better than the analogous calculation assuming
Poissonian arrivals. This will be the subject of the discussion presented 
in Section~\ref{sec:simulations}.

\item The distribution of the random delays added to each
arrival time has a very small impact on the distribution of the queue.
The only relevant parameter is the variance of the random delays. 
In order to have
a reliable forecast of the traffic over a congested hub one can use the
simplest distribution, e.g.\ the uniform one. This point will be discussed in
details in Section~\ref{sec:rob}.
\end{enumerate}

\section{Description of the arrival process}
\label{sec:arrivals}

The \textit{Pre-Scheduled Random Arrivals} (PSRA) process is
defined as follows.
Denoting  with  $\frac{1}{\lambda}$ the expected interarrival time between two
consecutive aircrafts, 
the actual arrival time of the $i$-th aircraft ($t_i\in\mathbb{R}$) is defined by
\begin{equation}
  \label{eq:1}
  t_i=\frac{i}{\lambda} +\xi_i \qquad i\in\mathbb{Z}
\end{equation}
where $\xi_i$'s are i.i.d.\ continuous random variables with
probability density $f^{(\sigma)}_\xi(t)$ and variance $\sigma^{2}$. 
Without loss of generality we can assume $\mathbb{E}(\xi_i)=0$, as
$\mathbb{E}(\xi_i)\ne 0$ affects only the initial configuration of the
system.

When $\sigma$ is large the process defined in~\eqref{eq:1} weakly
converges to the Poisson process, in particular it is possible to prove that 
its generating function tends pointwise to the
generating function of the Poisson process~\citep{gns}. 
This property also holds for a variant of the PSRA process that  takes into account the possibility of flights' cancellation 
as in \cite{ball1}. This variant is  an independent thinning version of this process, i.e. a process in which
each arrival has an independent probability $1-\gamma$ to be cancelled
(and the complementary probability $\gamma$ to be a \textit{true}
arrival).
In \textit{Air Traffic Management} (ATM) contexts it is natural to couple
a PSRA arrival process with a deterministic service process with expected
service rate $\lambda$. In this case the traffic
intensity of the thinned process is clearly $\rho = \gamma$.
  
Although both the PSRA process and its thinned version are 
very similar to the Poisson process
for large $\sigma$, they present a crucial difference with the latter: 
as soon as $\sigma$ remains finite, the PSRA process is negatively
autocorrelated.  
The covariance $\mathrm{Cov}(n_{1},n_{2})$ between the number of arrivals at two consecutive time periods  $n_{1}$ and $n_{2}$,
where $n_{1}$ is the number of arrivals in $(t,t+T]$ and $n_{2}$ is the number of arrivals in $(t+T,t+2T]$, is given by 
\[\mathrm{Cov}(n_{1},n_{2})=\mathbb{E}(n_{1}n_{2}) - \mathbb{E}
(n_{1})\,\mathbb{E} (n_{2}) = -\sum_{i} p_{i}^{(\sigma)}(t,t+T)
p_{i}^{(\sigma)}(t+T,t+2T) \]
where $p_{i}^{(\sigma)}(t_1,t_2)$ is the probability that the $i$-th aircraft
arrives in the interval $[t_1,t_2]$; for details, see~\citep{gns}.
A negative covariance means that $n_{1}$ and $n_{2}$ are
inversely correlated, thus a congested time slot is likely to be followed
or preceded by a slot with less-than-expected arrivals. 
Moreover, this proves that the hypothesis of
independence for $n_{1}$ and $n_{2}$, numbers of arrivals in different time slots,
is not correct, unless we are in the limit $\sigma \rightarrow \infty$. 

Simulations show that if we neglect this correlation
and we try to describe a queueing system 
with independent interarrival times, we obtain a gross overestimate for the average
length of the queue. This error is particularly rough when the system is congested (traffic
intensity $\rho$ near to~$1$), see Figure~\ref{fig:heathrow-fit} below.

\section{Real data from London Heathrow airport}
\label{sec:heathrow}

\begin{figure}[htbp]
  \centering
  \includegraphics[width=\textwidth]{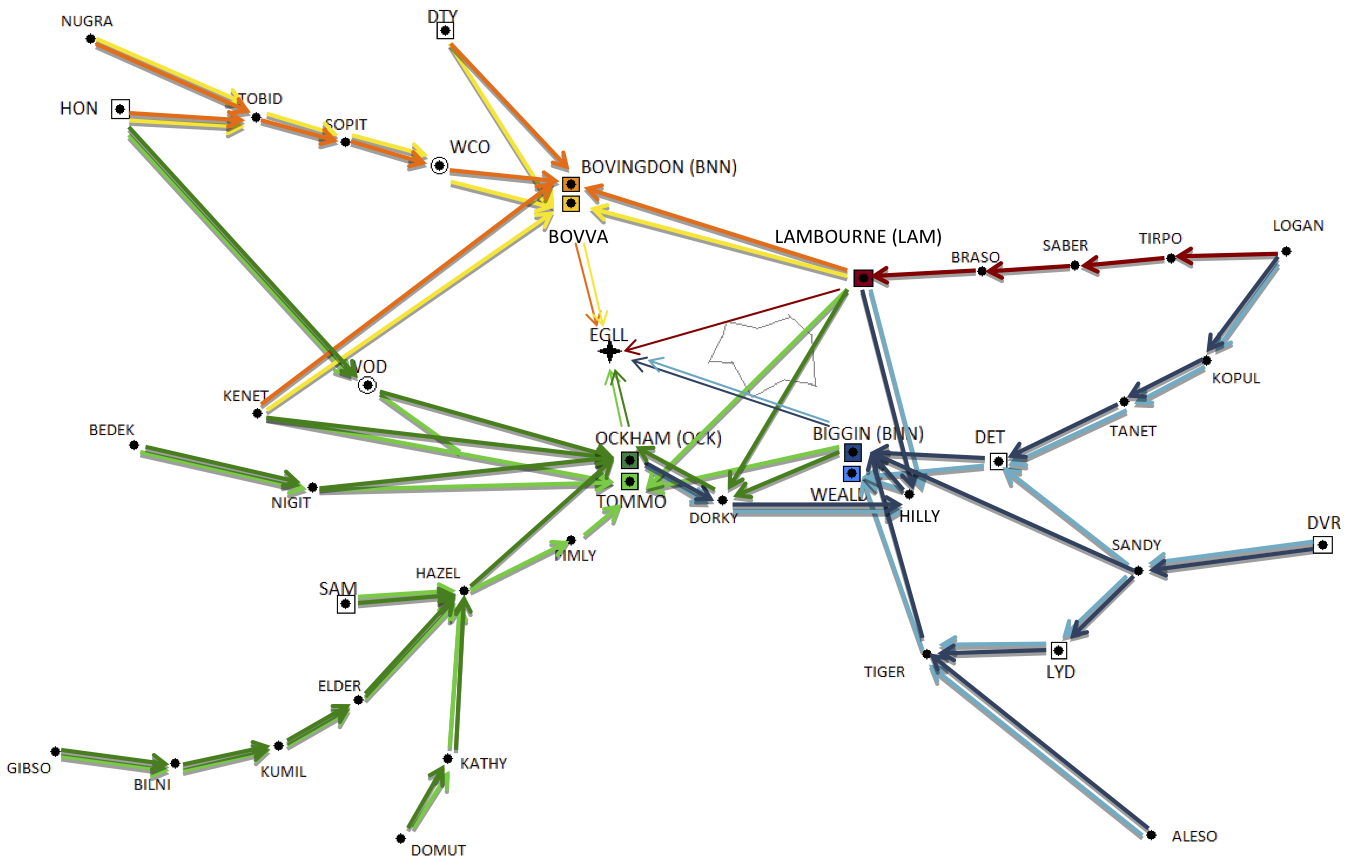}
  \caption{A sketch of the 28 STAR of the London Heathrow Airport.}
  \label{fig:arrivals}
\end{figure}

The London  Heathrow airport is served by two parallel independent runways.
The runways are used in a separate mode, meaning that  one runway is used only for departures and the other is used only for arrivals.
Aircraft occasionally are cleared to land on the departure runway to minimise delay subject to certain delay criteria.
According to the ICAO Standard Arrival Chart, Heathrow has 28 STAR
(standard arrival route) starting from 11 entry points; see Figure~\ref{fig:arrivals}.

\begin{figure}[htbp]
  \centering
  \includegraphics[width=\textwidth]{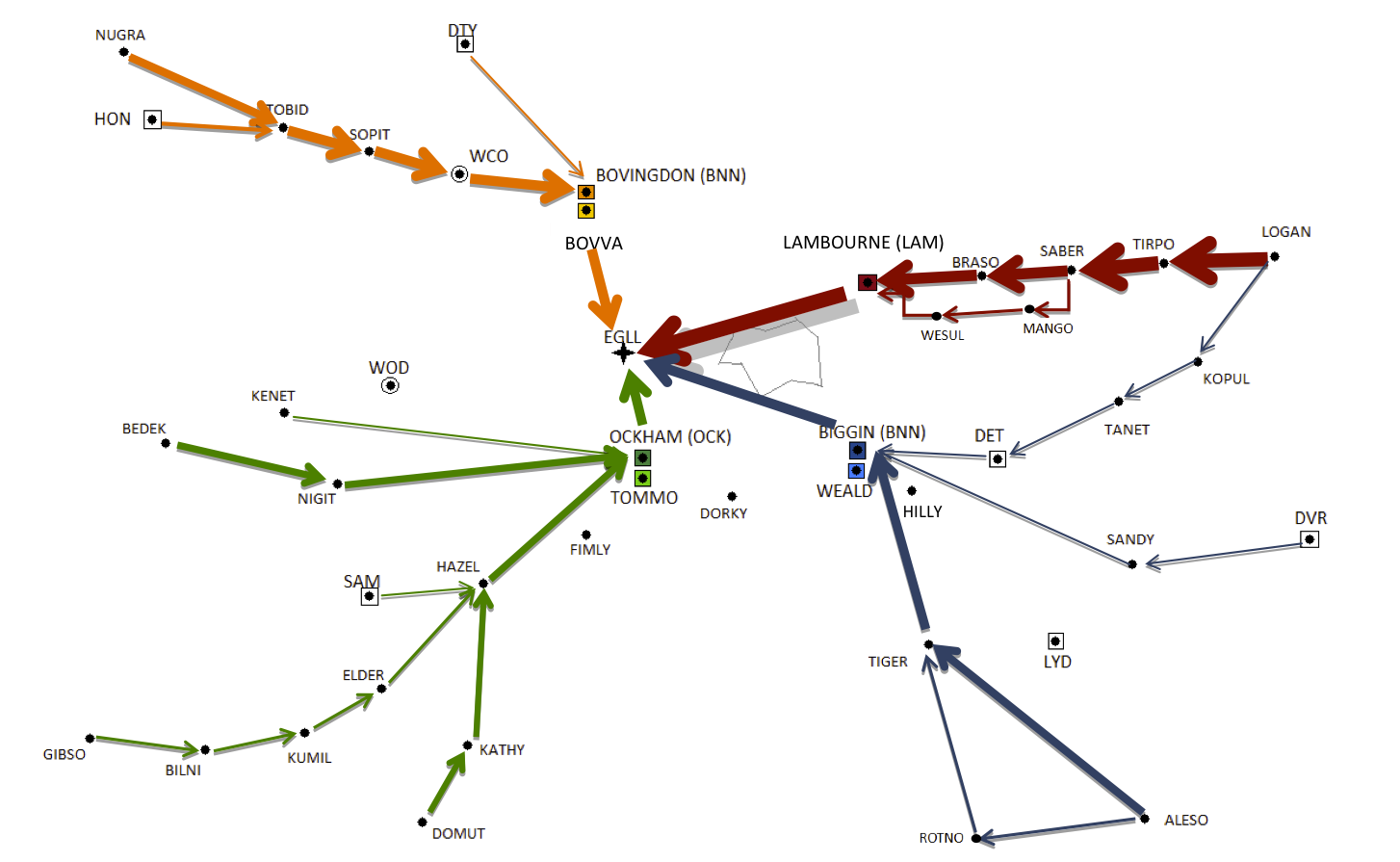}
  \caption{Qualitative layout of the incoming air traffic over the
    London Heathrow Airport during the time lapse under examination.}
  \label{fig:flux}
\end{figure}

In the intent of this analysis, actual data are used ranging from July
20 to July 30, 2010. This big database include 7,140 flights. A first
analysis of these data shows that most of the flights (about 70\%) enter
from three entry points: LOGAN, ALESO and NUGRA and pass on STAR LAM
3A, BIG 3B and BNN 1B, respectively. 
Figure~\ref{fig:flux} shows a qualitative layout
of the incoming air traffic in the aforesaid lapse of time. 

We limited ourselves to the three entry points mentioned above, 
and starting from this restricted database we observed that the
traffic intensity of the system was not completely constant during the
day but in the two slots from 6:00 a.m.\ to 10:30 a.m.\ and from 4:00
p.m.\ to 8:00 p.m.
For each flight landing in one of these two time intervals the
entrance time in the system has been subtracted to the landing time.
Let us call this difference \textit{approaching time}.
To obtain the amount of time each flight has been
waiting in queue, we subtract from the approaching time of any aircraft entering from a
given entry point the minimun approaching time from the
same entry point.
Note that the service time is nearly deterministic. This implies that 
the length of the queue that a given aircraft had to wait is proportional
to the time spent in queue by that aircraft.
Hence, the empirical
distribution of the time spent in queue is equivalent to that of the
queue lenght.
Eventually, we have gathered the data from the three distinct queues
and we have normalized the resulting histogram, which can be seen in
Figure~\ref{fig:heathrow-fit}.
All in all, the final number of flights considered was~4139. 


From the data we could also see that in the period above the average
number of flights entering in the airport fly 
area per hour is 40, while  the maximum number of flights landing on a
single runway is 41 per hour.
Therefore, in this analysis we will take the following measured value for 
the traffic intensity $\rho$,
\[\rho = \frac{\text{average number of arrivals per
    hour}}{\text{maximum number of landings per hour}} = \frac{40}{41} =
0.976\]

\section{Simulations of PSRA and comparison with Heathrow data}
\label{sec:simulations}

\begin{figure}[htbp]
  \centering
  \includegraphics[width=\textwidth]{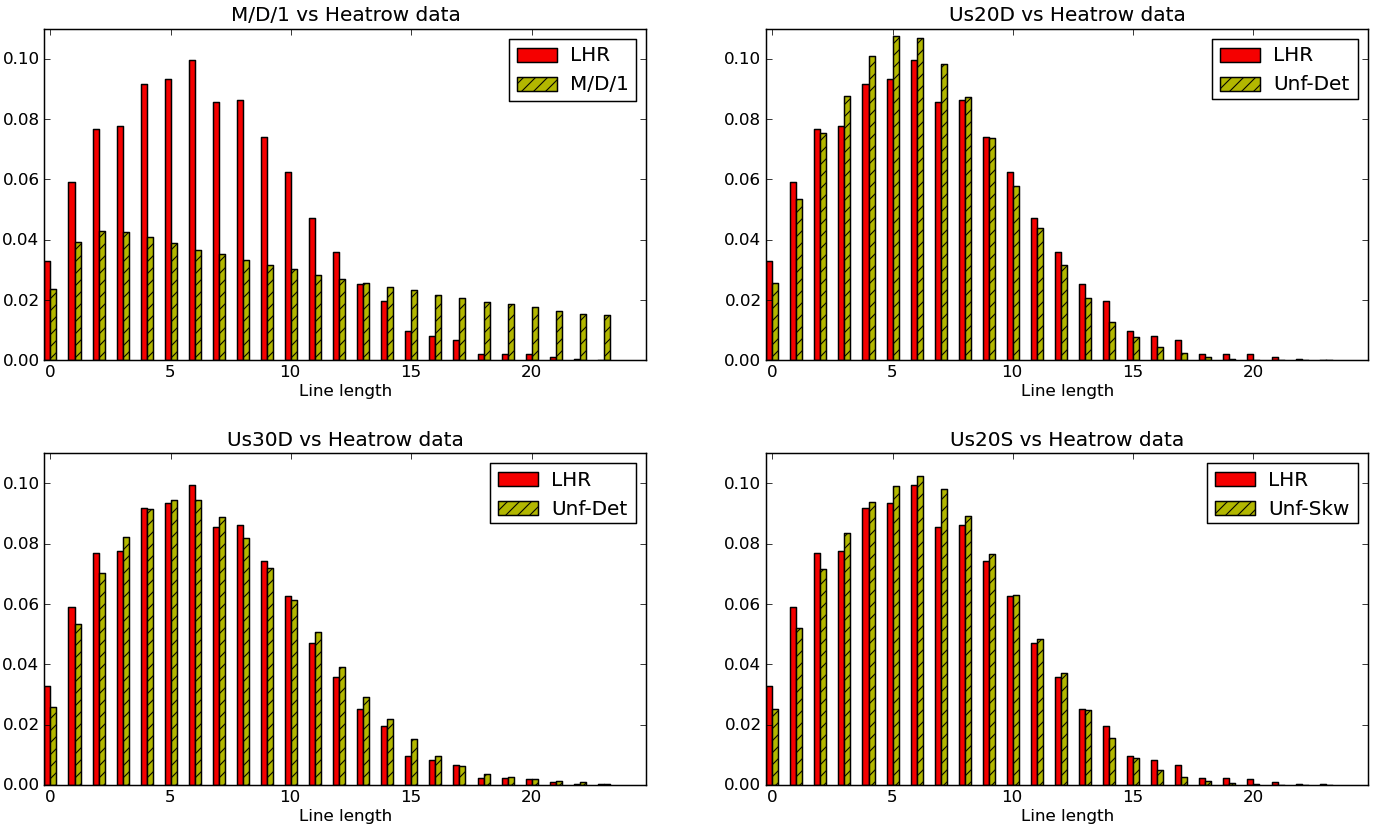}
  \caption{Fit of empirical distribution from Heathrow data with the
    empirical distribution from simulations of M/D/1 model and some
    variants of the PSRA model. 
    Subfigures \textit{a} to \textit{d} are displayed clockwise starting top left.} 
  \label{fig:heathrow-fit}
\end{figure}

The data discussed in Section~\ref{sec:heathrow} may be used to get an
idea of the high reliability of the PSRA model.
In Figure~\ref{fig:heathrow-fit}\textit{a}--\ref{fig:heathrow-fit}\textit{d} 
the empirical distribution derived
from actual data is compared with the empirical
distribution obtained by simulation of models of interest.
Subfigures \textit{a} to \textit{d} are displayed clockwise starting top left.

Figure~\ref{fig:heathrow-fit}\textit{a} refers to the M/D/1 case. Poissonian arrivals
are de facto a standard assumption in many works and studies in the
ATM field~\citep{balakrishnan2006scheduling, bauerle2007waiting,
  dunlay1976stochastic, marianov2003location, 
  willemain2004statistical} but they give not a satisfactory prediction 
of the tail of the queue distribution whatsoever.
Figure~\ref{fig:heathrow-fit}\textit{b} 
and~\ref{fig:heathrow-fit}\textit{d} refer to the PSRA with 
$\sigma = \frac{20}{\lambda}$ and $\sigma=\frac{30}{\lambda}$,
 respectively. These values are reasonable guesses of the order of the actual 
 standard deviation. We would need more data to validate this claim. 
Anyway, we clearly see that the PSRA model gives a very good
 accordance with empirical data, but to go one better and match what
 actually happens in real life operations we tried to introduce
 some randomness on the service time, modeled as a triangular random variable
 with mean $\frac{1}{\lambda}$ and mode $\frac{0.8}{\lambda}$.
The introduction of a such \textit{small} source of randomness 
 is completely legitimate by the need of the service providers to
 adjust the sequence of the arrivals on demand.
The variant of PSRA obtained this way is shown in Figure~\ref{fig:heathrow-fit}\textit{c}. 
We want to outline here that the simulations have been performed using a
straightforward
Python code. This is an important
features of this model: it can be easily simulated.

In Table~\ref{tab:distances} we show a measure of the accordance of the observed data with
the simulated ones using Total Variation and Hellinger distance. Note that the distance in the case
of memoryless arrivals is an order of magnitude larger than the distances obtained by PSRA arrivals.
\begin{savenotes}
  \begin{table}[htbp]
    \centering
    \begin{tabular}{lcc}
      \toprule
      Model & Total Variation distance & Hellinger distance\\
      \midrule
      M/D/1 & 0.41067 & 0.43903\\
      Uniform Delays\footnote{$\sigma=\nicefrac{20}{\lambda}$ and deterministic service time} & 0.07516 & 0.08133\\
      Uniform Delays\footnote{$\sigma=\nicefrac{30}{\lambda}$ and deterministic service time} & 0.03938 & 0.04565\\
      Uniform Delays\footnote{$\sigma=\nicefrac{20}{\lambda}$ and service times modeled as triangular random variable with mean $\nicefrac{1}{\lambda}$ and mode $\nicefrac{0.8}{\lambda}$ } & 0.05723 & 0.05814\\
      \bottomrule
    \end{tabular}
    \caption{Total Variation and Hellinger distances between the empirical distribution of Heathrow data and the distributions
      considered in Figure~\ref{fig:heathrow-fit} }
    \label{tab:distances}
  \end{table}
\end{savenotes}

\section{Robustness of the result}
\label{sec:rob}

Simulations were also used to study the robustness of the PSRA with
respect to the choice of the probability law $f_\xi$ of the delays. 
We considered delays of different type, namely \emph{uniform}, \emph{triangular},
\emph{normal} and \emph{exponential}, with zero mean and fixed
standard deviation $\sigma$.

\begin{figure}[!htb]
  \centering
  \includegraphics[width=\textwidth]{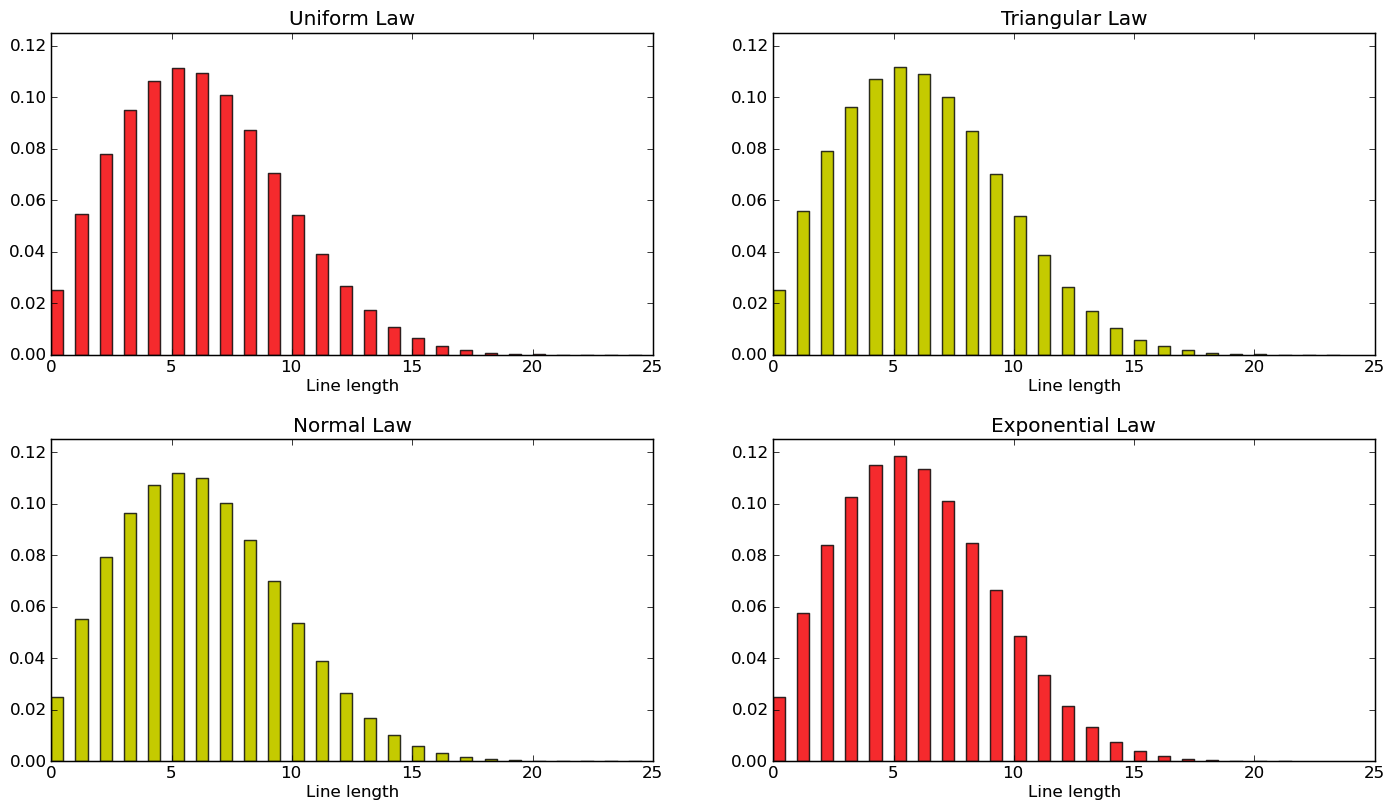}
  \caption{Robustness of the PSRA model with respect to the choice of the
    probability law $f_\xi$ of the delays. The delays have zero mean
    and fixed standard deviation
    $\sigma=\frac{20}{\lambda}$. Clockwise from top left histogram we
    have uniformly, triangularly, exponentially, and normally
    distributed $\xi$'s.}
  \label{fig:utne}
\end{figure}

Figure~\ref{fig:utne} suggests that the standard deviation of the delay
is the actual parameter of the model and that the resulting
queue distribution is insensitive of the nature of the delays.
For symmetrical probability laws such as uniform, triangular and
normal the histograms in Figure~\ref{fig:utne} present an astonishing
resemblance and differ for only small fluctuations.

Even more surprisingly, we obtain the same layout with a skew probability
law, like the exponential distribution. Figure~\ref{fig:utne} clearly
shows that the queue distribution obtained using exponentially delayed
arrivals is qualitatively the same as the uniformly delayed case.
This circumstance is of extreme interest for applications, for the
\textit{exact} solution of the model can be obtained for the case of exponential
delays. In~\cite{eda-2012} a PSRA model with exponentially distributed
$\xi$'s is considered and an iterative method to derive the explicit
expression of the generating function is provided. 
In the same paper it can be seen via simulations of the model that
the empirical and theoretical distributions completely agree. 

\section{Conclusions}
\label{sec:conc}
In this paper, we present a model to describe the inbound air traffic
over a congested hub. The model can be also used to study the possible
effects of the application of future technologies to the Air Traffic
Control~\citep{Iovanella2011}. 

This work is a preliminary attempt to study this important problem,
as such it should motivate further studies about the PSRA and its
performance analysis. Hopefully, the features of this arrival process
may give also an insight in the equally crucial problem of congestion
reduction in complex systems.


    \newpage
    \addtocontents{toc}{\protect\vspace{\beforebibskip}}
    \addcontentsline{toc}{section}{\refname}    
    \bibliographystyle{siam}
    \bibliography{cills}
\end{document}